\documentclass{ws-procs9x6-cpt16}

\usepackage{xspace} 

\begin{document}

\newcommand{\refeq}[1]{(\ref{#1})}
\def\etal {{\it et al.}}

\def\dzero  {\mbox{D0}\xspace}
\def\B       {{\ensuremath{B}}\xspace}
\def\Bbar    {{\ensuremath{\kern 0.18em\overline{\kern -0.18em B}{}}}\xspace}
\def\Bd      {{\ensuremath{\B^0}}\xspace}
\def\Bs      {{\ensuremath{\B^0_s}}\xspace}
\def\Bsb     {{\ensuremath{\Bbar{}^0_s}}\xspace}
\def\Bdb     {{\ensuremath{\Bbar{}^0}}\xspace}
\def\jpsi    {{\ensuremath{{J\mskip -3mu/\mskip -2mu\psi\mskip 2mu}}}\xspace}
\def\Kbar    {{\kern 0.2em\overline{\kern -0.2em K}{}}\xspace}
\def\Kz      {{\ensuremath{K^0}}\xspace}
\def\Kzb     {{\ensuremath{\Kbar{}^0}}\xspace}
\def\KS      {{\ensuremath{K^0_{\mathrm{ \scriptscriptstyle S}}}}\xspace}
\def\Kp      {{\ensuremath{K^+}}\xspace}
\def\Km      {{\ensuremath{K^-}}\xspace}
\def\Dbar    {{\kern 0.2em\overline{\kern -0.2em D}{}}\xspace}
\def\D       {{\ensuremath{D}}\xspace}
\def\Db      {{\ensuremath{\Dbar}}\xspace}
\def\Dz      {{\ensuremath{\D^0}}\xspace}
\def\Dzb     {{\ensuremath{\Dbar{}^0}}\xspace}
\def\Dsmp    {{\ensuremath{D^{\mp}_s}}\xspace}
\def\pip     {{\ensuremath{\pi^+}}\xspace}
\def\invps   {\ensuremath{{\mathrm{ \,ps^{-1}}}}\xspace}
\def\invfb   {\ensuremath{\mbox{\,fb}^{-1}}\xspace}
\newcommand{\tev}{\ensuremath{\mathrm{\,Te\kern -0.1em V}}\xspace}
\newcommand{\gev}{\ensuremath{\mathrm{\,Ge\kern -0.1em V}}\xspace}
\def\hr   {\ensuremath{\mathrm{ \,hr}}\xspace}
\newcommand{\dm}{{\ensuremath{\Delta m}}\xspace}
\newcommand{\dmd}{{\ensuremath{\Delta m_d}}\xspace}
\newcommand{\DG}{{\ensuremath{\Delta\Gamma}}\xspace}
\newcommand{\DGd}{{\ensuremath{\Delta\Gamma_d}}\xspace}
\newcommand{\Rez}{\ensuremath{\mathcal{R}e(z)}\xspace}
\newcommand{\Imz}{\ensuremath{\mathcal{I}m(z)}\xspace}
\newcommand{\mean}[1]{\ensuremath{\left\langle #1 \right\rangle}} 
\def\Dan          {{\ensuremath{\Delta a_0}}\xspace}
\def\Dax          {{\ensuremath{\Delta a_X}}\xspace}
\def\Day          {{\ensuremath{\Delta a_Y}}\xspace}
\def\Daz          {{\ensuremath{\Delta a_Z}}\xspace}
\def\Damu         {{\ensuremath{\Delta a_{\mu}}}\xspace}
\def\Dap          {{\ensuremath{\Delta a_{\parallel}}}\xspace}
\def\Dao          {{\ensuremath{\Delta a_{\perp}}}\xspace}
\def\DamuUnit {\ensuremath{\times 10^{-15}\gev}}
\def\DasUnit  {\ensuremath{\times 10^{-13}\gev}}
\def\DamuBsUnit {\ensuremath{\times 10^{-14}\gev}}
\def\DasBsUnit  {\ensuremath{\times 10^{-12}\gev}}

\title{Measurements of CPT Violation at LHCb}

\author{J.\ van Tilburg}

\address{Nikhef, Science Park 105\\ 
1098 XG Amsterdam, Netherlands}

\author{On behalf of the LHCb Collaboration}

\begin{abstract}
Recent measurements of CPT violation and Lorentz symmetry breaking in $\Bd-\Bdb$
mixing and $\Bs-\Bsb$ mixing, obtained from data taken by the LHCb experiment,
are highlighted. The results are expressed in terms of the Standard-Model
Extension (SME) coefficients, which incorporate both CPT and Lorentz
violation. Due to the large boost of the \B mesons at LHCb, 
the SME coefficients can be determined with high precision. 
The bounds on these coefficients are improved significantly compared 
to previous measurements.
\end{abstract}

\bodymatter

\section{Introduction}
\label{sec:intro}

The LHCb detector\cite{Alves:2008zz, LHCb-DP-2014-002} is a single-arm forward
spectrometer designed for the study of heavy flavor hadrons. Many results have
been published by the LHCb collaboration, in particular on CP violation in
decays of $b$ and $c$ hadrons. In contrast, until recently LHCb had made no
measurements on CPT violation in these decays. In these proceedings, a new
result\cite{LHCb-PAPER-2016-005} from the LHCb collaboration on CPT violation in
$\Bd-\Bdb$ mixing and $\Bs-\Bsb$ mixing is highlighted.

Violation of CPT symmetry implies a breaking of Lorentz invariance in a local,
interacting quantum field theory.\cite{Greenberg:2002uu} This means that any
CPT-violating parameter must also violate Lorentz invariance. The Standard-Model
Extension (SME) is an effective field theory, where CPT- and Lorentz-violating
terms are added to the Standard-Model lagrangian.\cite{Colladay:1996iz,
 Colladay:1998fq} This framework provides the experimental opportunity to
measure the coupling coefficients in these terms. The LHCb
result\cite{LHCb-PAPER-2016-005} presented here is given in terms of these SME
coefficients.

In the past, there have been many experimental searches for CPT violation in
neutral-meson systems.\cite{PDG2014,vanTilburg:2014dka} The majority of these
searches have been done without any assumption on the breaking of Lorentz
invariance, referred to as the classical approach. In the last 15 years, more
searches have been performed within the SME framework, placing tight constraints
on its coefficients.\cite{Kostelecky:2008ts}

\section{Formalism of CPT violation in neutral-meson systems}

The particle-antiparticle mixing between neutral-meson states creates an
interferometric system that enhances the sensitivity to CPT violation
enormously. Conservation of CPT symmetry implies equal mass and lifetime of
particles and antiparticles. The CPT-violating observable in the mixing process
is given by
\begin{align}
 z = \frac{\delta m - i\delta\Gamma/2}{\dm + i\DG/2} \ ,
\end{align}
where $\delta m$ and $\delta\Gamma$ are the (CPT-violating) mass and decay
width differences between the particle and antiparticle states. The high
sensitivity to $z$ comes through the small values in the denominator of the
eigenvalue differences, \dm and \DG, of the two-state system. In the SME
framework, the $z$ observable becomes\cite{Kostelecky:1997mh,
 Kostelecky:2001ff}
\begin{align}
 z = \frac{\beta^{\mu}\Damu}{\dm + i\DG/2} \ ,
 \label{eq:z}
\end{align}
where $\beta^{\mu}=(\gamma,\gamma\vec{\beta})$ is the four velocity of the
neutral meson and \Damu is a real four-vector vacuum expectation value that
describes the coupling with the mesons. The complex parameter $z$ can be
determined directly from the decay rates as function of the decay time of the
neutral meson.\cite{vanTilburg:2014dka, LHCb-PAPER-2016-005}

There are four systems of neutral mesons. In all of them, the mixing formalism
is identical, however, their phenomenology is very different owing to the
different values of \dm and \DG, and number of decay modes. In the $\Kz-\Kzb$
system, there have been many searches for CPT violation by dedicated kaon
experiments (KLOE, KTeV, CPLEAR, and NA48) following the classical approach. An
experimental overview is given in Ref.\ \refcite{PDG2014}. Strong constraints on
the SME coefficients have been made using data from KLOE, KTeV and
E773.\cite{Kostelecky:2008ts} It will be difficult for LHCb to compete with
these dedicated kaon experiments due to the lower statistics and worse kaon
lifetime acceptance. The situation is already different in the $\Dz-\Dzb$
system. Only a single measurement exists, by the FOCUS
collaboration,\cite{Link:2002fg} using a sample of 35k $\Dz\to\Km\pip$
decays. LHCb would be able to improve this measurement significantly owing to
the 50M $\Dz\to\Km\pip$ decays, collected during Run 1.\cite{vanTilburg:2014dka}
In the following, I will focus on the two remaining neutral-meson systems: the
\Bd and \Bs systems.

\section{Measurements at LHCb}

In both the \Bd system and \Bs system, \DG is negligibly small compared to
\dm. The Standard Model predicts that \DG is about a factor 200 smaller than
\dm,\cite{Artuso:2015swg} which is already confirmed in the \Bs system.  Since
\Damu is real, it follows from Eq.\ \refeq{eq:z} that \Imz is a factor 400
smaller than \Rez. Therefore, to constrain the SME coefficients, \B decays to CP
eigenstates are used, which are more sensitive to \Rez in comparison to using \B
decays to flavor-specific final states.\cite{vanTilburg:2014dka} The golden \B
decay modes to CP eigenstates, $\Bd\to\jpsi\KS$ and $\Bs\to\jpsi\Kp\Km$, have
been used due to their relatively large branching fraction. These modes have
been studied already at LHCb to measure $\sin(2\beta)$ and
$\phi_s$.\cite{LHCb-PAPER-2015-004,LHCb-PAPER-2014-059} For the present
analysis, the fit models have been extended to allow for possible CPT
violation.\cite{LHCb-PAPER-2016-005} The results\cite{LHCb-PAPER-2016-005} are
shown in Table \ref{tab:CPTresults}. No significant sidereal variation and no
violation of CPT symmetry are observed. In the \Bd system, there is a large
improvement of three orders of magnitude with respect to the previous best
result\cite{Aubert:2007bp} from BaBar.  In the \Bs system, there is an order of
magnitude improvement with respect to the previous best
result\cite{Abazov:2015ana} from \dzero. The improvements are primarily
attributed to the large boost of the \B mesons at LHCb (i.e.,
$\mean{\beta\gamma} \approx 20$ versus $\mean{\beta\gamma} = 0.5$ at BaBar and
$\mean{\beta\gamma} = 4.7$ at \dzero).

\begin{table}
 \tbl{Results on \Damu for the decay channels $\Bd\to\jpsi\KS$ and
   $\Bs\to\jpsi\Kp\Km$.}
   {\begin{tabular}{r@{} @{}l r@{} @{}l}\toprule
 \multicolumn{2}{c}{\Bd system} & \multicolumn{2}{c}{\Bs system} \\ \colrule
 \Dap &= $(-0.10 \pm 0.82 \pm 0.54) \DamuUnit$ &
 \Dap &=$(-0.89 \pm 1.41 \pm 0.36)\DamuBsUnit$ \\
 \Dao &= $(-0.20 \pm 0.22 \pm 0.04) \DasUnit$ &
 \Dao &=$(-0.48 \pm 0.39 \pm 0.08)\DasBsUnit$ \\
 \Dax &= $(+1.97 \pm 1.30 \pm 0.29) \DamuUnit$ &
 \Dax &=$(+1.01 \pm 2.08 \pm 0.71)\DamuBsUnit$ \\
 \Day &= $(+0.44 \pm 1.26 \pm 0.29) \DamuUnit$ &
 \Day &=$(-3.83 \pm 2.09 \pm 0.71)\DamuBsUnit$ \\
& & \Rez &= $-0.022 \pm 0.033 \pm 0.003$ \\
& & \Imz &= $\phantom{+}0.004 \pm 0.011 \pm 0.002$ \\ \botrule
   \end{tabular}}
   \label{tab:CPTresults}
\end{table}

\section{Summary and outlook}

In summary, interferometry with neutral mesons provides a sensitive method to
test violations of CPT symmetry and Lorentz invariance. The LHCb experiment is
well suited to improve the SME bounds, in particular due to the high boost of
the particles produced at the LHC. There are plans to measure $z$ and \Damu in
the \Dz system, which aim to improve the current bounds by a factor 40. As
highlighted in these proceedings, greatly improved limits on CPT violation and
Lorentz symmetry breaking in \B mixing have been published by
LHCb.\cite{LHCb-PAPER-2016-005} These results are based on an integrated
luminosity of 3\invfb obtained in Run 1 of the LHC. At the end of Run 2 in 2019,
an expected $4$-$6\invfb$ will be added. Due to the larger cross sections at the
new center-of-mass energy of 13\tev, the heavy flavor yields are almost a
factor two higher in Run 2. Furthermore, the \B meson boost will also be about
30\% higher. Together this means that the uncertainties will reduce by a factor
two. A further improvement can be expected from the upgraded LHCb detector that
will start data taking after 2019: with a projected 50\invfb the uncertainties
will drop by more than a factor of six.

\section*{Acknowledgments}

This work is supported by the Netherlands Organisation for Scientific Research
(NWO Vidi grant 680-47-523) and the Foundation for Fundamental Research on
Matter (FOM).

\end{document}